\RequirePackage{fix-cm}
\documentclass[smallextended]{svjour3}       % onecolumn (second format)
\smartqed  % flush right qed marks, e.g. at end of proof
\usepackage{graphicx,amssymb,amsfonts,mathrsfs,latexsym,amsmath,amssymb}
\journalname{xxx}
\begin{document}

\title{The Gray image of constacyclic codes over
the finite chain ring $F_{p^m}[u]/\langle u^k\rangle$%\thanks{Grants or other notes
} \subtitle{}

\titlerunning{The Gray image of constacyclic codes over
the finite chain ring $F_{p^m}[u]/\langle u^k\rangle$}

\author{ Yuan Cao \and Yonglin Cao
         %etc.
}


\institute{Yonglin Cao (corresponding author) \at
             \email{ylcao@sdut.edu.cn}\\
           Yuan Cao\at
             \email{woodwest2@163.com}\\
           School of Sciences, Shandong University of Technology, Zibo, Shandong 255091, China                %  \\
}

\date{Received: date / Accepted: date}

\maketitle

\begin{abstract}
Let $\mathbb{F}_{p^m}$ be a finite field of
cardinality $p^m$, where $p$ is a prime, and $k, N$ be any positive integers. We denote $R_k=F_{p^m}[u]/\langle u^k\rangle
=F_{p^m}+uF_{p^m}+\ldots+u^{k-1}F_{p^m}$ ($u^k=0$) and
$\lambda=a_0+a_1u+\ldots+a_{k-1}u^{k-1}$ where $a_0, a_1,\ldots, a_{k-1}\in F_{p^m}$ satisfying
$a_0\neq 0$ and $a_1=1$. Let $r$ be a positive integer satisfying $p^{r-1}+1\leq k\leq p^r$. First we define a Gray map from $R_k$ to
$F_{p^m}^{p^r}$ first, then
prove that the Gray image of any linear $\lambda$-constacyclic code over
$R_k$ of length $N$ is a distance preserving linear $a_0^{p^r}$-constacyclic code over $F_{p^m}$ of length $p^rN$.
Furthermore, the
generator polynomials for each linear $\lambda$-constacyclic code over
$R_k$ of length $N$ and its Gray image are given respectively. Finally, some optimal
constacyclic codes over $F_{3}$ and $F_{5}$ are constructed.

\end{abstract}

\section{Introduction}
\noindent
  Algebraic coding theory deals with the design of error-correcting and error-detecting codes for reliable transmission
of information across noisy channel. The class of constacyclic codes plays a very significant role in
the theory of error-correcting codes as they can be efficiently encoded with simple shift
registers. This family of codes is thus interesting for both theoretical and practical reasons.

\par
  Let $\Gamma$ be a commutative finite ring with identity $1\neq 0$, and $\Gamma^{\times}$ be the multiplicative group of invertible elements of
$\Gamma$. For any $a\in
\Gamma$, we denote by $\langle a\rangle_\Gamma$, or $\langle a\rangle$ for
simplicity, the ideal of $\Gamma$ generated by $a$, i.e., $\langle
a\rangle_\Gamma=a\Gamma=\{ab\mid b\in \Gamma\}$. For any ideal $I$ of $\Gamma$, we will identify the
element $a+I$ of the residue class ring $\Gamma/I$ with $a$ (mod $I$) for
any $a\in \Gamma$ in this paper.

\par
   A \textit{code} over $\Gamma$ of length $N$ is a nonempty subset ${\cal C}$ of $\Gamma^N=\{(a_0,a_1,\ldots$, $a_{N-1})\mid a_j\in\Gamma, \
j=0,1,\ldots,N-1\}$. The code ${\cal C}$
is said to be \textit{linear} if ${\cal C}$ is an $\Gamma$-submodule of $\Gamma^N$. All codes in this paper are assumed to be linear.
   Let $\gamma\in \Gamma^{\times}$.
Then a linear code
${\cal C}$ over $\Gamma$ of length $N$ is
called a $\gamma$-\textit{constacyclic code}
if $(\gamma c_{N-1},c_0,c_1,\ldots,c_{N-2})\in {\cal C}$ for all
$(c_0,c_1,\ldots,c_{N-1})\in{\cal C}$. Particularly, ${\cal C}$ is
called a \textit{negacyclic code} if $\gamma=-1$, and ${\cal C}$  a  \textit{cyclic code} if $\gamma=1$.
  For any $a=(a_0,a_1,\ldots,a_{N-1})\in \Gamma^N$, let
$a(x)=a_0+a_1x+\ldots+a_{N-1}x^{N-1}\in \Gamma[x]/\langle x^N-\gamma\rangle$. We will identify $a$ with $a(x)$ in
this paper. By Dinh [5] Propositions 2.2, we have

\vskip 3mm \noindent
  {\bf Lemma 1.1}  \textit{Let $\gamma\in \Gamma^{\times}$. Then ${\cal C}$ is a  $\gamma$-constacyclic code
of length $N$ over $\Gamma$ if and only if ${\cal C}$ is an ideal of
the residue class ring $\Gamma[x]/\langle x^N-\gamma\rangle$}.

\vskip 3mm \par
   In 1994, Hammons et al. [7] discovered that some efficient nonlinear binary codes can be viewed
as the Gray images of some linear codes over $\mathbb{Z}_4$. From then on, the construction of the Gray map over finite rings has been a topic of
study. In [14], Wolfmann showed that Gray images of a linear single-root negacyclic code over $\mathbb{Z}_4$
is a distance invariant binary cyclic code (not necessarily linear). Tapia-Recillas and Vega in [13]
obtained that the Gray image of a $1+2^k$-constacyclic code over $\mathbb{Z}_{2^{k+1}}$ was a binary distance
invariant quasi-cyclic code. Later, the result was generalized to $\mathbb{Z}_{p^{k+1}}$ by Ling and Blackford in [9].

\par
   Qian et al. in [10] proved that the Gray image of a linear single-root $(1+u)$-constacyclic code over $F_2+uF_2$
is a binary distance invariant linear cyclic code. Amarra and Nemenzo. in [2] discussed the Gray image of
a single-root $(1-u)$-constacyclic code over $F_{p^k}+uF_{p^k}$, which was a quasi-cyclic code over $F_p^k$.
Abular and Siap in [1] showed that the Gray image of a linear $(1+u)$-constacyclic code of an
arbitrary length over $F_2+uF_2$ was a binary distance invariant linear cyclic code and got the
generator polynomial of the Gray image. Sobhani and Esmaeili in [12] studied the Gray image of a
single-root $(1+u^t)$-constacyclic code over $F_q[u]/\langle u^{t+1}\rangle$, which was a quasi-cyclic code over $F_q$.
Kai et al. in [8] showed that the Gray image of a linear $(1+\lambda u)$-constacyclic code with an
arbitrary length over $F_{p}[u]/\langle u^k\rangle$ was a distance invariant linear code over $F_{p}$.
Cao [3] determined generator polynomials
of $(1+w\gamma)$-constacyclic codes and their dual codes over an arbitrary finite
chain ring $R$, where $w$ is a unit of $R$ and $\gamma$ generates the unique maximal ideal of $R$.
The Gray image of a
single-root $(1+u+u^2)$-constacyclic code over the ring $F_2[u]/\langle u^3\rangle$ was proved to be a binary
distance invariant linear cyclic code in [11]. In [4], Ding and Li defined a new Gray map from $R_k=F_{2^m}[u]/\langle u^k\rangle$
to $F_{2^m}^{2^j}$, where $2^{j-1}+1\leq k\leq 2^j$ for some positive integer $j$, and studied the Gray images of
$(1+u+\ldots+u^{k-1})$-constacyclic codes over $R_k$ via the Gray map. At the end of the paper, a natural problem is proposed:
How can we extend the result to the
ring $F_{p^m}[u]/\langle u^k\rangle$? The main result and their proofs in [4] depends  on $1+u+\ldots+u^{k-1}$ and the characteristic $2$
of the finite field $F_{2^m}$, so we will adopt a new method to solve this problem.

\par
  In this paper, we adopt the following notations.

\vskip 3mm \noindent
  {\bf Notation 1.2} Let $F_{p^m}$ be a finite field of cardinality $p^m$, where
$p$ is a prime and $m$ be a positive integer, $a_0, a_2,.., a_{k-1}$ be fixed elements
of $F_{p^m}$ satisfying $a_0\neq 0$, and $k, r$ be integers satisfying
$2\leq p^{r-1}+1\leq k\leq p^r$.
We denote

\vskip 2mm \par
   $\bullet$ $R_k=F_{p^m}[u]/\langle u^k\rangle
=F_{p^m}+u F_{p^m}+\ldots+u^{k-1}F_{p^m} \ (u^k=0)$.

\vskip 2mm \par
   $\bullet$ $N=p^en$ where $e$ is a nonnegative integer and $n$ is a positive integer satisfying ${\rm gcd}(p,n)=1$.

\vskip 2mm \par
   $\bullet$ $\lambda=a_0+u+a_2u^2+\ldots+a_{k-1}u^{k-1}\in R_k^{\times}$.

\vskip 2mm \par
   $\bullet$ $\vartheta=a_0^{p^r}\in F_{p^m}^{\times}$ and $\omega=x^N-a_0\in F_{p^m}[x]$.

\vskip 2mm \par
   $\bullet$ $a(u)=u+a_2u^2+\ldots+a_{k-1}u^{k-1}\in \mathbb{F}_{p^m}[u]$.

\vskip 2mm \par
   $\bullet$ $J_k=\left(\begin{array}{cc} 0 & 0\cr I_{k-1} & 0\end{array}\right)\in
{\rm M}_k(F_{p^m})$, where
$I_i$ is the identity matrix of order $i$ for any positive integer $i$.

\vskip 2mm \par
   $\bullet$ $a(J_k)=J_k+\sum_{i=2}^{k-1}a_iJ_k^i
=\left(\begin{array}{ccccc}0    &    &    &    &  \cr
                           1    &  0 &    &    &  \cr
                           a_2  &  1 &  0 &    &  \cr
                           \vdots & \ddots & \ddots & \ddots & \cr
                           a_{k-1}& \ldots & a_2 & 1 & 0 \end{array}\right)\in {\rm M}_k(F_{p^m}).$

\vskip 3mm \par
  This paper is
organized as follows. Section 2 presents properties of matrices $J_k$ and $a(J_k)$.
In Section 3, we define a Gray map from $R_k$ to $F_{p^m}^{p^r}$,
and construct a distance-preserving map $\Phi$ from $(R_k[x]/\langle x^N-\lambda\rangle$, Lee distance) to
($F_{p^m}[x]/\langle x^{p^rN}-\vartheta\rangle$, Hamming distance). In
Section 4, we prove that $\Phi$ is an injective $F_{p^m}[x]$-module homomorphism with image
${\rm Im}(\Phi)=\langle (x^N-a_0)^{p^r-k}\rangle$. Then $\Phi$ induces
a bijection between $\lambda$-constacyclic codes over $R_k$ of length $N$
and $\vartheta$-constacyclic codes over $F_{p^m}$ of length $p^rN$ contained in $\langle (x^N-a_0)^{p^r-k}\rangle$. Moreover,
we provide generator polynomials for each of these two kind of constacyclic codes.
Finally, some optimal linear constacyclic
codes over $F_3$ and $F_5$ are constructed under the Gray map in Section 5.

%%%%%%%%%%%%%%%%%%%%%%%%%%%%%%%%%%%%%%%%%%%%%%%%%%%%%%%%%%%%%%%%%%%%%%%

%%%%%%%%%%%%%%%%%%%%%%%%%%%%%%%%%%%%%%%%%%%%%%%%%%%%%%%%%%%%%%%%%%%%%%

\section{Preliminaries}
\noindent
   In this section, we provide an algorithm to find a fixed invertible matrix
$P_k\in {\rm M}_k(F_{p^m})$ satisfying $P_ka(J_k)=J_kP_k$. We will use
$P_k$ to define a Gray map from $R_k$ to $F_{p^m}^{p^r}$ and build a relation between $\lambda$-constacyclic codes over $R_k$ of length $N$
and $\vartheta$-constacyclic codes over $F_{p^m}$ of length $p^rN$.
From now on, the transpose of a matrix $B$ will be denoted by
$B^{{\rm tr}}$.

\vskip 3mm \noindent
 {\bf Theorem 2.1}
  \textit{An invertible matrix $P_k\in {\rm M}_k(F_{p^m})$ satisfying $P_ka(J_k)=J_kP_k$,
can be constructed by the following recursion formula}:

\noindent
 $\bullet$  $P_2=I_2$;

\noindent
 $\bullet$ \textit{When $i\geq 3$, let $P_i=\left(\begin{array}{cc} 1 & 0\cr Y_i & P_{i-1}\end{array}\right)$, where $Y_i=(y_1^{(i)},\ldots,y_{i-2}^{(i)},y_{i-1}^{(i)})^{{\rm tr}}$ satisfying}
$$(1,y_1^{(i)},\ldots,y_{i-2}^{(i)})^{{\rm tr}}=P_{i-1}(1,a_2,\ldots,a_{i-1})^{{\rm tr}}
\ {\rm and} \ y_{i-1}^{(i)}=0.$$

\vskip 3mm \noindent
  {\bf Proof.}  Let $k\geq 2$. As $P_2=I_2$, we see that $P_2$ is invertible and satisfies $P_2a(J_2)=J_2P_2$.

\par
  Assume $k\geq 3$ and that $P_{k-1}$ is an invertible matrix in ${\rm M}_{k-1}(F_{p^m})$ satisfying $P_{k-1}a(J_{k-1})=J_{k-1}P_{k-1}$. Now,
let $P_{k}=\left(\begin{array}{cc} 1 & 0\cr Y_{k} & P_{k-1}\end{array}\right)$, where $Y_{k}=(y_1^{(k)},\ldots,y_{k-2}^{(k)},0)^{{\rm tr}}$ satisfying
\begin{equation}
(1,y_1^{(k)},\ldots,y_{k-2}^{(k)})^{{\rm tr}}=P_{k-1}(1,a_2,\ldots,a_{k-1})^{{\rm tr}}.
\end{equation}
It is obvious that $P_k$ is invertible by the mathematical induction. Denote
$$e_1=(1,0,\ldots,0)^{{\rm tr}} \ {\rm and} \ \xi=(1,a_2,\ldots,a_{k-1})^{{\rm tr}}.$$
Then we have $J_k=\left(\begin{array}{cc} 0 & 0\cr e_1 & J_{k-1}\end{array}\right)$
and $a(J_k)=
\left(\begin{array}{cc} 0 & 0\cr \xi & a(J_{k-1})\end{array}\right)$ in which $a(J_{k-1}) =J_{k-1}+\sum_{i=2}^{k-1}a_iJ_{k-1}^i=J_{k-1}+\sum_{i=2}^{k-2}a_iJ_{k-1}^i$ as $J_{k-1}^{k-1}=0$.
Therefore,
\begin{equation}
J_kP_k=\left(\begin{array}{cc} 0 & 0\cr e_1 & J_{k-1}\end{array}\right)\left(\begin{array}{cc} 1 & 0\cr Y_{k} & P_{k-1}\end{array}\right)
=\left(\begin{array}{cc} 0 & 0\cr e_1+J_{k-1}Y_{k} & J_{k-1}P_{k-1}\end{array}\right)
\end{equation}
where
\begin{equation}
e_1+J_{k-1}Y_{k}=(1,y_1^{(k)},\ldots,y_{k-2}^{(k)})^{{\rm tr}}=P_{k-1}(1,a_2,\ldots,a_{k-1})^{{\rm tr}}=P_{k-1}\xi
\end{equation}
by Equation (1). On the other hand, we have
\begin{equation}
P_ka(J_k)=\left(\begin{array}{cc} 1 & 0\cr Y_{k} & P_{k-1}\end{array}\right)\left(\begin{array}{cc} 0 & 0\cr \xi & a(J_{k-1})\end{array}\right)
=\left(\begin{array}{cc} 0 & 0\cr P_{k-1}\xi & P_{k-1}a(J_{k-1})\end{array}\right).
\end{equation}
Since, by $P_{k-1}a(J_{k-1})=J_{k-1}P_{k-1}$ and by Equations (2)--(4) we conclude
that $P_{k}a(J_{k})=J_{k}P_{k}$.
\hfill $\Box$

\vskip 3mm \noindent
   {\bf Corollary 2.2} \textit{Using the notations above, we have the following conclusions}:

\vskip 2mm\par
   (i) \textit{$(1,u,\ldots,u^{k-1})a(u)=a(u)(1,u,\ldots,u^{k-1})=(1,u,\ldots,u^{k-1})a(J_k)$}.

\vskip 2mm\par
   (ii) \textit{$(1,(x^N-a_0),(x^N-a_0)^2,\ldots,(x^N-a_0)^{k-1})=(1,u,\ldots,u^{k-1})P_k^{-1}Y$
in the ring $R_k[x]/\langle x^N-\lambda\rangle$, where}
 $$Y=\left(\begin{array}{cccc} 1 & & & \cr y_1^{(k)} & 1 & & \cr \vdots &\ddots &\ddots &
\cr y_{k-1}^{(k)} & \ldots & y_1^{(k)} & 1\end{array}\right)=I_k+\sum_{i=1}^{k-1}y_i^{(k)}J_k^i$$
 \textit{being an invertible matrix in ${\rm M}_{k\times k}(F_{p^m})$}.

\vskip 3mm \noindent
   {\bf Proof.} (i) It follows from $u(1,u,\ldots,u^{k-1})=(1,u,\ldots,u^{k-1})J_k$.

\par
   (ii) In the ring $R_k[x]/\langle x^N-\lambda\rangle$,
we have $x^N=\lambda=a_0+u+\sum_{i=2}^{k-1}a_iu^{i}$, which implies that that
$$(x^N-a_0)=a(u)=(1,u,u^2,\ldots,u^{k-1})a(J_k)\varepsilon_1, $$
where $e_1=(1,0,\ldots,0)^{{\rm tr}}$ and $a(J_k)=P_k^{-1}J_kP_k$ by Theorem 2.1. From this and by (i), using a recursive method we deduce that
\begin{eqnarray*}
(x^N-a_0)^{j}&=&a(u)^{j-1}a(u)=\left(a(u)^{j-1}(1,u,u^2,\ldots,u^{k-1})\right)a(J_k)e_1\\
 &=&\left((1,u,u^2,\ldots,u^{k-1})a(J_k)^{j-1}\right)a(J_k)e_1\\
 &=&(1,u,u^2,\ldots,u^{k-1})a(J_k)^je_1\\
 &=&(1,u,u^2,\ldots,u^{k-1})P_k^{-1}J_k^jP_ke_1
\end{eqnarray*}
for all $j=1,2,\ldots,k-1$. By Theorem 2.1, we know that $P_{k}=\left(\begin{array}{cc} 1 & 0\cr Y_{k} & P_{k-1}\end{array}\right)$, where $Y_{k}=(y_1^{(k)},\ldots,y_{k-2}^{(k)},0)^{{\rm tr}}$, which implies that
$$J_k^jP_ke_1=J_k^j(P_ke_1)=J_k^j\left(\begin{array}{c} 1 \cr Y_{k}\end{array}\right)
\left(0,\ldots,0,1,y_1^{(k)},\ldots,y_{k-j-1}^{(k)}\right)^{{\rm tr}}$$
for all $j=1,2,\ldots,k-1$. Therefore, we obtain
\begin{eqnarray*}
&&\left(1,(x^N-a_0),(x^N-a_0)^2,\ldots,(x^N-a_0)^{k-1}\right)\\
&=&(1,u,\ldots,u^{k-1})\left(P_k^{-1}(P_ke_1),P_k^{-1}J_kP_ke_1,P_k^{-1}J_k^2P_ke_1,
\ldots,P_k^{-1}J_k^{k-1}P_ke_1\right)\\
&=&(1,u,u^2,\ldots,u^{k-1})\left(P_k^{-1}(P_ke_1,J_kP_ke_1,J_k^2P_ke_1,
\ldots,J_k^{k-1}P_ke_1)\right)\\
&=&(1,u,u^2,\ldots,u^{k-1})P_k^{-1}Y,
\end{eqnarray*}
where $Y=\left(P_ke_1,J_kP_ke_1,J_k^2P_ke_1,
\ldots,J_k^{k-1}P_ke_1\right)$ as required.
\hfill $\Box$

 \vskip 3mm \noindent
 {\bf Example 2.3}

\par
   (i) Let $R_3=F_3[u]/\langle u^3\rangle$ and $\lambda=2+u+2u^2$. $P_3=\left(\begin{array}{ccc} 1 & 0 & 0\cr 2 & 1 & 0 \cr 0 & 0 & 1\end{array}\right)$.

\par
   (i) Let $R_ 4=F_5[u]/\langle u^4\rangle$ and $\lambda=2+u+2u^2+4u^3$. $P_4=\left(\begin{array}{cccc} 1 & 0 & 0 & 0\cr 4 & 1 & 0 & 0 \cr 4 & 2 & 1 & 0 \cr 0 & 0 & 0 & 1\end{array}\right)$.

%%%%%%%%%%%%%%%%%%%%%%%%%%%%%%%%%%%%%%%%%%%%%%%%%%%%%%%%%%%%%%%%%%%%%%
%%%%%%%%%%%%%%%%%%%%%%%%%%%%%%%%%%%%%%%%%%%%%%%%%%%%%%%%%%%%%%%%%%%%%%%

%%%%%%%%%%%%%%%%%%%%%%%%%%%%%%%%%%%%%%%%%%%%%%%%%%%%%%%%%%%%%%%%%%%%%%%%%%%%%%%%%%%%%%%%%

%%%%%%%%%%%%%%%%%%%%%%%%%%%%%%%%%%%%%%%%%%%%%%%%%%%%%%%%%%%%%%%%%%%%%%%%%%%%%%%%%%%%%%%%%%%%%%
\section{A Gray map from $R_k$ to $F_{p^m}^{p^r}$ and a Lee weight on $R_k[x]/\langle x^N-\lambda\rangle$} \label{}
\noindent
   In this section, we define a Gray map from $R_k$ to $F_{p^m}^{p^r}$ using the matrix
$P_k$ constructed by Theorem 2.1. By use of this Gray map and the Hamming weight on
$F_{p^m}^{p^r}$ we define
Lee weights and Lee distances for elements of $R_k^N$. In order to
investigate $\lambda$-constacyclic codes over $R_k$ of length $N$,
i.e., ideals of the ring $R_k[x]/\langle x^N-\lambda\rangle$, we construct a distance-preserving map $\Phi$ from $(R_k[x]/\langle x^N-\lambda\rangle$, Lee distance) to
$(F_{p^m}[x]/\langle x^{p^rN}-\vartheta\rangle$, Hamming distance).

\par
  Let $F_{p^m}[z]/\langle (z-a_0)^{p^r}\rangle$ be the the residue class ring
of the polynomial ring $F_{p^m}[z]$ module its ideal $\langle (z-a_0)^{p^r}\rangle$
generated by $(z-a_0)^{p^r}$. Then each element $f(z)$ of $F_{p^m}[z]/\langle (z-a_0)^{p^r}\rangle$
can be uniquely
expressed as $f(z)=\sum_{j=0}^{p^r-1}f_jz^j$ where $f_0,f_1,\ldots,f_{p^r-1}\in F_{p^m}$. It is
well known that
$$f(z)\mapsto \underline{v}_f=(f_0,f_1,\ldots,f_{p^r-1})$$
is an $F_{p^m}$-linear isomorphism from
$F_{p^m}[z]/\langle (z-a_0)^{p^r}\rangle$ onto $F_{p^m}^{p^r}$.
In the rest of the paper, we will identify $F_{p^m}[z]/\langle (z-a_0)^{p^r}\rangle$ with $F_{p^m}^{p^r}$
under this $F_{p^m}$-linear isomorphism. As usual, the
Hamming weight of ${\rm wt}_H(f(z))$ is defined by
$${\rm wt}_H(f(z))
={\rm wt}_H(\underline{v}_f)=|\{j\mid f_j\neq 0, \ 0\leq j\leq p^r-1\}|.$$

\vskip 1mm \noindent
   {\bf Definition 3.1} Let $\alpha=b_0+b_1u+\ldots+b_{k-1}u^{k-1}=(1,u,\ldots,u^{k-1})B\in R_k$,
where $B=(b_0,b_1,\ldots,b_{k-1})^{{\rm tr}}$ with $b_i\in F_{p^m}$ for all $i=0,1,\dots,k-1$.

\par
  $\bullet$ We define
a \textit{Gray map} $\varphi:R_k\rightarrow F_{p^m}[z]/\langle (z-a_0)^{p^r}\rangle$ by
\begin{eqnarray*}
\varphi(\alpha)&=&((z-a_0)^{p^r-k},(z-a_0)^{p^r-k+1},\ldots,(z-a_0)^{p^r-1})(P_kB)\\
  &=&h_0(z-a_0)^{p^r-k}+h_1(z-a_0)^{p^r-k+1}+\ldots+h_{k-1}(z-a_0)^{p^r-1},
\end{eqnarray*}
where $(h_0,h_1,\ldots,h_{k-1})^{{\rm tr}}=P_kB\in {\rm M}_{k\times 1}(F_{p^m})$.

\par
  $\bullet$ Using the Gray map $\varphi$, we define the \textit{Lee weight} of $\alpha$ by
$${\rm wt}_L(\alpha)={\rm wt}_H(\varphi(\alpha)),$$
where ${\rm wt}_H(\varphi(\alpha))$ is the Hamming weight of $\varphi(\alpha)\in F_{p^m}[z]/\langle (z-a_0)^{p^r}\rangle$.

\par
  $\bullet$ For any $\alpha_1,\alpha_2\in R_k$ the \textit{Lee distance} of $\alpha_1$ and $\alpha_2$
is defined as the Lee weight ${\rm wt}_L(\alpha_1-\alpha_2)$ of $\alpha_1-\alpha_2$.

\vskip 3mm \noindent
  {\bf Remark} It is clear that $\varphi$ is injective since $P_k$ is invertible. As $F_{p^m}[z]/\langle (z-a_0)^{p^r}\rangle$ has been identified with $F_{p^m}^{p^r}$, $\varphi$ is a Gray map from $R_k$ to $F_{p^m}^{p^r}$.

\vskip 3mm\par
   Then we extend the Gray map $\varphi$ to $R_k^N$ in the natural way.

\vskip 3mm \noindent
   {\bf Definition 3.2} Let $\xi=(\alpha_0,\alpha_1,\ldots,\alpha_{N-1})\in R_k^N$,
where $\alpha_j\in R_k$ for all $j=0,1,\dots,N-1$.

\par
  $\bullet$ Define
a map $\widetilde{\varphi}: R_k^N\rightarrow \left(F_{p^m}[z]/\langle (z-a_0)^{p^r}\rangle\right)^N$ by
$$\widetilde{\varphi}(\xi)=\widetilde{\varphi}(\alpha_0,\alpha_1,\ldots,\alpha_{N-1})=(\varphi(\alpha_0),\varphi(\alpha_1),\ldots,
\varphi(\alpha_{N-1})).$$

\par
  $\bullet$  Define the \textit{Lee weight} of $\xi$ by
$${\rm wt}_L(\xi)=\sum_{j=0}^{N-1}{\rm wt}_L(\alpha_j)=\sum_{j=0}^{N-1}{\rm wt}_H(\varphi(\alpha_j))={\rm wt}_H(\widetilde{\varphi}(\xi)).$$

\par
  $\bullet$  Define the \textit{Lee distance} of $\xi,\eta\in R_k^N$ as the Lee weight ${\rm wt}_L(\xi-\eta)$
of $\xi-\eta\in R_k^N$.

\vskip 3mm \par
   The following lemma follows from the definitions of $\widetilde{\varphi}$ and Lee
weight immediately.

\vskip 3mm \noindent
   {\bf Lemma 3.3} \textit{The map $\widetilde{\varphi}$ defined by Definition 3.2 is an injective $F_{p^m}$-linear homomorphism from
$R_k^N$ to $\left(F_{p^m}[z]/\langle (z-a_0)^{p^r}\rangle\right)^N$ and a distance-preserving map from $(R_k^N$, Lee distance$)$ to
$((F_{p^m}[z]/\langle (z-a_0)^{p^r}\rangle)^N$, Hamming distance$)$}.

\vskip 3mm \noindent
  {\bf Definition 3.4} Let $\beta=(b_0(z),b_1(z),\ldots,b_{N-1}(z))\in
\left(F_{p^m}[z]/\langle (z-a_0)^{p^r}\rangle\right)^N$ where $b_j(x)\in F_{p^m}[z]/\langle (z-a_0)^{p^r}\rangle$
for all $j=0,1,\ldots,N-1$.

\par
  $\bullet$ Define a map $\sigma: \left(F_{p^m}[z]/\langle (z-a_0)^{p^r}\rangle\right)^N\rightarrow F_{p^m}[x]/\langle x^{p^rN}-\vartheta\rangle$ by
\begin{eqnarray*}
\sigma(\beta)&=&\left(b_0(x^N),b_1(x^N),\ldots,b_{N-1}(x^N)\right)(1,x,\ldots,x^{N-1})^{{\rm tr}}\\
&=&\sum_{j=0}^{N-1}x^jb_j(x^N).
\end{eqnarray*}

\vskip 3mm \par
  It is obvious that $\sigma$ is well-defined and injective. Moreover, we have

\vskip 3mm \noindent
   {\bf Lemma 3.5} \textit{The map $\sigma$ defined by Definition 3.4 is an injective $F_{p^m}$-linear homomorphism and a distance-preserving map from $((F_{p^m}[z]/\langle (z-a_0)^{p^r}\rangle)^N$,  Hamming distance$)$ to
$(F_{p^m}[x]/\langle x^{p^rN}-\vartheta\rangle$, Hamming distance$)$}.

\vskip 3mm \noindent
   {\bf Proof.} By Definition 3.4, one can easily verify that $\sigma$ is an an injective $F_{p^m}$-linear homomorphism.
For any $\beta=(b_0(z),b_1(z),\ldots,b_{N-1}(z))\in
\left(F_{p^m}[z]/\langle (z-a_0)^{p^r}\rangle\right)^N$, where $b_j(x)\in F_{p^m}[z]/\langle (z-a_0)^{p^r}\rangle$
for all $j=0,1,\ldots$, $N-1$, it is clear that
$${\rm wt}_H(\beta)=\sum_{j=0}^{N-1}{\rm wt}_H((b_j(z))
={\rm wt}_H(\sum_{j=0}^{N-1}x^jb_j(x^N))
={\rm wt}_H(\sigma(\beta)).$$
Hence $\sigma$ is a Hamming distance-preserving map.
\hfill $\Box$

\vskip 3mm \par
   Now, we define a map $\tau: R_k[x]/\langle x^N-\lambda\rangle\rightarrow R_k^N$ by
$$\tau(\alpha_0+\alpha_1x+\ldots+\alpha_{N-1}x^{N-1})=(\alpha_0,\alpha_1,\ldots,\alpha_{N-1})$$
for all $\alpha_0,\alpha_1,\ldots,\alpha_{N-1}\in R_k$. It is clear that $\tau$ is an
$R_k$-module isomorphism from $R_k[x]/\langle x^N-\lambda\rangle$ onto $R_k^N$. As usual, the Lee weight
of $\alpha_0+\alpha_1x+\ldots+\alpha_{N-1}x^{N-1}$ is defined as
$${\rm wt}_L(\alpha_0+\alpha_1x+\ldots+\alpha_{N-1}x^{N-1})={\rm wt}_L(\alpha_0,\alpha_1,\ldots,\alpha_{N-1})=\sum_{j=0}^{N-1}{\rm wt}_L(\alpha_j).$$
Then it is obvious that $\tau$ is a Lee distance-preserving map.

\par
  Finally, let $\Phi=\sigma\widetilde{\varphi}\tau$. Then by Definitions 3.2 and 3.4, we have the following commutative diagram:
$$\left.\begin{array}{ccc} R_k[x]/\langle x^N-\lambda\rangle & \stackrel{\Phi}{\longrightarrow} & F_{p^m}[x]/\langle x^{p^rN}-\vartheta\rangle
\cr \tau \ \downarrow &  & \uparrow \ \sigma \cr R_k^N & \stackrel{\widetilde{\varphi}}{\longrightarrow} & (F_{p^m}[z]/\langle (z-a_0)^{p^r}\rangle)^N.\end{array}\right.$$
As stated above, by Lemmas 3.3 and 3.5 we deduce the following Theorem.

\vskip 3mm \noindent
   {\bf Theorem 3.6} \textit{The map $\Phi$ defined above is injective
and a distance-preserving map from $(R_k[x]/\langle x^N-\lambda\rangle$, Lee distance$)$ to
$(F_{p^m}[x]/\langle x^{p^rN}-\vartheta\rangle$, Hamming distance$)$}.

\vskip 3mm \noindent
   {\bf Remark} By the commutative diagram above, we see that the map $\Phi$ is completely determined
by the Gray map $\varphi: R_k\rightarrow F_{p^m}[z]/\langle (z-a_0)^{p^r}\rangle$ defined in Definition 3.1.

%%%%%%%%%%%%%%%%%%%%%%%%%%%%%%%%%%%%%%%%%%%%%%%%%%%%%%%%%%%%%%%%%%%%%%

\section{ Relationship between $\lambda$-constacyclic codes over $R_k$ of length $N$
and $\vartheta$-constacyclic codes over $F_{p^m}$ of length $p^rN$}
\noindent
In this section, we investigate the algebraic properties of the map $\Phi$ defined
in Section 3. Then we establish a bijection between $\lambda$-constacyclic codes over $R_k$ of length $N$
and a subset of $\vartheta$-constacyclic codes over $F_{p^m}$ of length $p^rN$ using $\Phi$. To do this, we give a clear expression
for $\Phi$ first.

\par
  As $R_k=F_{p^m}+u F_{p^m}+\ldots+u^{k-1}F_{p^m}$ ($u^k=0$),
each element $c(x)$ of $R_k[x]/\langle x^N-\lambda\rangle$ can be uniquely
expressed as $c(x)=\sum_{j=0}^{N-1}(\sum_{i=0}^{k-1}c_{i,j}u^i)x^j$ with $c_{i,j}\in F_{p^m}$.
Hence $c(x)$ is uniquely written as a product of matrixes:
\begin{equation}
c(x)=(1,u,\ldots,u^{k-1})A_{c(x)}(1,x,\ldots,x^{N-1})^{{\rm tr}},
\end{equation}
where $A_{c(x)}=\left(\begin{array}{cccc}c_{0,0} & c_{0,1} & \ldots & c_{0,N-1}
\cr c_{1,0} & c_{1,1} & \ldots & c_{1,N-1} \cr \ldots & \ldots & \ldots & \ldots
\cr c_{k-1,0} & c_{k-1,1} & \ldots & c_{k-1,N-1}\end{array}\right)\in {\rm M}_{k\times N}(F_{p^m})$.

\vskip 3mm \noindent
   {\bf Lemma 4.1} Using the representation (5) for each $c(x)\in R_k[x]/\langle x^N-\lambda\rangle$,
the map $\Phi$ from $R_k[x]/\langle x^N-\lambda\rangle$ to $F_{p^m}[x]/\langle x^{p^rN}-\vartheta\rangle$
is given by:
$$\Phi(c(x))=(\omega^{p^r-k},\omega^{p^r-k+1},\ldots,\omega^{p^r-1})(P_kA_{c(x)})(1,x,\ldots,x^{N-1})^{{\rm tr}},$$
where $\omega=x^N-a_0\in F_{p^m}[x]$.

\vskip 3mm \noindent
  {\bf Proof.} Let $c(x)\in R_k[x]/\langle x^N-\lambda\rangle$. Using the representation of $c(x)$ given by Equation (5), we write
$A_{c(x)}=(\eta_0,\eta_1,\ldots,\eta_{N-1})$, where $\eta_j$ is the $(j+1)$th column vector of the matrix
$A_{c(x)}$ for $0\leq j\leq N-1$. Then by (5) we have
$$c(x)=\alpha_0+\alpha_1x+\alpha_{N-1}x^{N-1} \ {\rm with} \ \alpha_j=(1,u,\ldots,u^{k-1})\eta_j,$$
which implies that $\tau(c(x))=(\alpha_0,\alpha_1,\ldots,\alpha_{N-1})\in R_k^N$.
For each $j=0,1,\ldots,N-1$, by Definition 3.1 it follows that
$$\varphi(\alpha_j)=((z-a_0)^{p^r-k},(z-a_0)^{p^r-k+1},\ldots,(z-a_0)^{p^r-1})(P_k\eta_j).$$
From this and by the definition
of $\widetilde{\varphi}$, we deduce
\begin{eqnarray*}
\widetilde{\varphi}(\tau(c(x)))&=&\widetilde{\varphi}(\alpha_0,\alpha_1,\ldots,\alpha_{N-1})
     =\left(\varphi(\alpha_0),\varphi(\alpha_1),\ldots,\varphi(\alpha_{N-1})\right)\\
    &=&((z-a_0)^{p^r-k},\ldots,(z-a_0)^{p^r-1})(P_k\eta_0,P_k\eta_1,\ldots,P_k\eta_{N-1})\\
   &=&((z-a_0)^{p^r-k},\ldots,(z-a_0)^{p^r-1})P_kA_{c(x)}.
\end{eqnarray*}
Then by $\Phi=\sigma\widetilde{\varphi}\tau$,  Definition 3.4 and Lemma 3.5 we obtain
\begin{eqnarray*}
\Phi(c(x))&=&\sigma(\widetilde{\varphi}(\tau(c(x))))\\
&=&((x^N-a_0)^{p^r-k},\ldots,(x^N-a_0)^{p^r-1})P_kA_{c(x)}(1,x,\ldots,x^{N-1})^{{\rm tr}}.
\end{eqnarray*}
\hfill $\Box$

\vskip 3mm
   As $F_{p^m}$ is a subfield of $R_k$, $F_{p^m}[x]$ is a subring of $R_k[x]$. For any $f(x)\in F_{p^m}[x]$, $\alpha(X)\in R_k[x]/\langle x^N-\lambda\rangle$ and $g(x)\in F_{p^m}[x]/\langle x^{p^rN}-\vartheta\rangle$, define
$$f(x)\circ \alpha(X)=f(x)\alpha(x) \ ({\rm mod} \ x^N-\lambda),$$
\begin{equation}
f(x)\circ g(X)=f(x)g(x) \ ({\rm mod} \ x^{p^rN}-v),
\end{equation}
respectively, where $f(x)\alpha(x)$ and $f(x)g(x)$ are usual products of polynomials in $F_{p^m}[x]$. Then both $R_k[x]/\langle x^N-\lambda\rangle$ and $F_{p^m}[x]/\langle x^{p^rN}-v\rangle$
are $F_{p^m}[x]$-modules. In order to simplify notations, we will write $f(x)\circ \alpha(X)$ and $f(x)\circ g(X)$
simply as $f(x)\alpha(x)$ and $f(x)g(x)$, respectively.

\vskip 3mm \noindent
   {\bf Lemma 4.2} \textit{Using the notations above, $\Phi$ is an injective $F_{p^m}[x]$-module homomorphism from $R_k[x]/\langle x^N-\lambda\rangle$ to $F_{p^m}[x]/\langle x^{p^rN}-\vartheta\rangle$ satisfying}
$${\rm Im}(\Phi)=\langle (x^N-a_0)^{p^r-k}\rangle$$
\textit{which is an ideal of $F_{p^m}[x]/\langle x^{p^rN}-\vartheta\rangle$ generated by $(x^N-a_0)^{p^r-k}$}.

\vskip 3mm \noindent
   {\bf Proof.} It is obvious that $\Phi$ is injective, since $P_k$
is an invertible matrix by Lemma 2.1. For any $c_1(x),c_2(x)\in R_k[x]/\langle x^N-\lambda\rangle$ and $b\in F_{p^m}$,
by Equation (5) and Lemma 4.1 it follows that $A_{bc_1(x)+c_2(x)}=bA_{c_1(x)}+A_{c_2(x)}$. Hence
\begin{eqnarray*}
\Phi(bc_1(x)+c_2(x))&=&(\omega^{p^r-k},\ldots,\omega^{p^r-1})
\left(P_kA_{bc_1(x)+c_2(x)}\right)(1,x,\ldots,x^{N-1})^{{\rm tr}}\\
&=&b(\omega^{p^r-k},\ldots,\omega^{p^r-1})
\left(P_kA_{c_1(x)}\right)(1,x,\ldots,x^{N-1})^{{\rm tr}}\\
&&+(\omega^{p^r-k},\ldots,\omega^{p^r-1})
\left(P_kA_{c_2(x)}\right)(1,x,\ldots,x^{N-1})^{{\rm tr}}\\
&=&b\Phi(c_1(x))+\Phi(c_2(x)).
\end{eqnarray*}
Hence $\Phi$ is an injective $F_{p^m}$-linear space homomorphism.

\par
   Let $c(x)\in R_k[x]/\langle x^N-\lambda\rangle$ and denote $\eta=(0,\ldots,0,1)^{{\rm tr}}$.
By Corollary 2.2(i), it follows that $(1,u,\ldots,u^{k-1})a(u)=(1,u,\ldots,u^{k-1})a(J_k)$.
Moreover, in the ring $R_k[x]/\langle x^N-\lambda\rangle$,
we have $x^N=\lambda=a_0+a(u)$. Then by Equation (5), we obtain
 \begin{eqnarray*}
 xc(x)&=&(1,u,\ldots,u^{k-1})A_{c(x)}\left(x(1,x,\ldots,x^{N-2},x^{N-1})^{{\rm tr}}\right)\\
 &=&(1,u,\ldots,u^{k-1})A_{c(x)}(x,x^2,\ldots,x^{N-1},x^N)^{{\rm tr}}\\
 &=&(1,u,\ldots,u^{k-1})A_{c(x)}(x,x^2,\ldots,x^{N-1},a_0+a(u))^{{\rm tr}}\\
 &=&(1,u,\ldots,u^{k-1})A_{c(x)}\left((x,x^2,\ldots,x^{N-1},a_0)^{{\rm tr}}+(0,\ldots,0,a(u))^{{\rm tr}}\right)\\
 &=&(1,u,\ldots,u^{k-1})A_{c(x)}(x,x^2,\ldots,x^{N-1},a_0)^{{\rm tr}}\\
   &&+(1,u,\ldots,u^{k-1})a(u)A_{c(x)}(0,\ldots,0,1)^{{\rm tr}}\\
 &=&(1,u,\ldots,u^{k-1})\left(A_{c(x)} \left(\begin{array}{cc} 0 & I_{N-1}\cr a_0 & 0\end{array}\right)\right)
    (1,x,x^2,\ldots,x^{N-1})^{{\rm tr}}\\
   &&+(1,u,\ldots,u^{k-1})(a(J_k)A_{c(x)})\eta.
\end{eqnarray*}
 From this and by Lemma 4.1, we deduce that
\begin{eqnarray*}
\Phi(xc(x))&=&(\omega^{p^r-k},\ldots,\omega^{p^r-1})\left(P_kA_{c(x)} \left(\begin{array}{cc} 0 & I_{N-1}\cr a_0 & 0\end{array}\right)\right)
    (1,x,\ldots,x^{N-1})^{{\rm tr}}\\
   &&+(\omega^{p^r-k},\ldots,\omega^{p^r-1})(P_ka(J)A_{c(x)})\eta.
\end{eqnarray*}
On the other hand, in the ring $F_{p^m}[x]/\langle x^{p^rN}-\vartheta\rangle$ we have
$\omega^{p^r}=(x^N-a_0)^{p^r}=x^{p^rN}-a_0^{p^r}=x^{p^rN}-\vartheta=0$, which implies that
$$(\omega^{p^r-k},\ldots,\omega^{p^r-1})\omega=(\omega^{p^r-k+1},\ldots,\omega^{p^r-1},0)=(\omega^{p^r-k},\ldots,\omega^{p^r-1})J_k.$$
Then by Lemma 4.1 and $x^N=a_0+(x^N-a_0)=a_0+\omega$, we have
\begin{eqnarray*}
x\Phi(c(x))&=&(\omega^{p^r-k},\ldots,\omega^{p^r-1})(P_kA_{c(x)})(x,\ldots,x^{N-1},x^N)^{{\rm tr}}\\
&=&(\omega^{p^r-k},\ldots,\omega^{p^r-1})(P_kA_{c(x)})(x,x^2,\ldots,x^{N-1},a_0)^{{\rm tr}}\\
   &&+(\omega^{p^r-k},\ldots,\omega^{p^r-1})\omega(P_kA_{c(x)})(0,\ldots,0,1)^{{\rm tr}}\\
 &=&(\omega^{p^r-k},\ldots,\omega^{p^r-1})\left(P_kA_{c(x)} \left(\begin{array}{cc} 0 & I_{N-1}\cr a_0 & 0\end{array}\right)\right)
    (1,x,\ldots,x^{N-1})^{{\rm tr}}\\
   &&+(\omega^{p^r-k},\ldots,\omega^{p^r-1})J_k(P_kA_{c(x)})\eta.
\end{eqnarray*}
Since $P_ka(J_k)=J_kP_k$ by Theorem 2.1, we have $\Phi(xc(x))=x\Phi(c(x))$.

\par
   As stated above, we conclude that
$\Phi$ is an injective $F_{p^m}[x]$-module homomorphism from $R_k[x]/\langle x^N-\lambda\rangle$ to $F_{p^m}[x]/\langle x^{p^rN}-\vartheta\rangle$.

\par
   By Lemma 4.1, Equation (5) and $\omega=x^N-a_0$ it
follows that
\begin{eqnarray*}
{\rm Im}(\Phi)&=&\{(\omega^{p^r-k},\omega^{p^r-k+1},\ldots,\omega^{p^r-1})(P_kA)
   \left(\begin{array}{c} 1 \cr x \cr \ldots \cr x^{N-1}\end{array}\right)
    \mid A\in {\rm M}_{k\times N}(F_{p^m})\}\\
    &=&\{\omega^{p^r-k}(1,\omega,\ldots,\omega^{k-1})B
   \left(\begin{array}{c} 1 \cr x \cr \ldots \cr x^{N-1}\end{array}\right)
    \mid B\in {\rm M}_{k\times N}(F_{p^m})\}\\
    &\subseteq&\langle (x^N-a_0)^{p^r-k}\rangle.
\end{eqnarray*}
Conversely, for any $g(x)\in \langle (x^N-a_0)^{p^r-k}\rangle$ there exists $b(x)\in F_{p^m}[x]/\langle x^{p^rN}-\vartheta\rangle$
such that $g(x)=(x^N-a_0)^{p^r-k}b(x)$ (mod $x^{p^rN}-\vartheta$). We regard $b(x)$ as a
polynomial in $F_{p^m}[x]$ having degree less than $p^rN$. Using division with a remainder, we obtain
$b_0(x),b_1(x),\ldots,b_{p^r-1}(x)\in F_{p^m}[x]$ both having degree less than $N$ such that
$b(x)=\sum_{i=0}^{p^r-1}(x^N-a_0)^ib_i(x)$. From this and by
$(x^N-a_0)^{p^r}=x^{p^rN}-a_0^{p^r}=x^{p^rN}-\vartheta=0$ in $F_{p^m}[x]/\langle x^{p^rN}-\vartheta\rangle$, we deduce
\begin{eqnarray*}
g(x)&=&(x^N-a_0)^{p^r-k}b_0(x)+(x^N-a_0)^{p^r-k+1}b_1(x)\\
    &&+\ldots+(x^N-a_0)^{p^r-1}b_{k-1}(x)\\
    &=&(\omega^{p^r-k},\omega^{p^r-k+1},\ldots,\omega^{p^r-1})B(1,x,\ldots,x^{N-1})^{{\rm tr}},
\end{eqnarray*}
where $B\in {\rm M}_{k\times N}(F_{p^m})$ satisfying
$$B(1,x,\ldots,x^{N-1})^{{\rm tr}}=(b_0(x),b_1(x),\ldots,b_{k-1}(x))^{{\rm tr}}.$$
Since $P_k$ is an invertible matrix by Theorem 2.1, we have that
$$c(x)=(1,u,\ldots,u^{k-1})(P_k^{-1}B)(1,x,\ldots,x^{N-1})^{{\rm tr}}\in R_k[x]/\langle x^N-\lambda\rangle$$
satisfying $\Phi(c(x))=g(x)$ by Lemma 4.1. Hence $\langle (x^N-a_0)^{p^r-k}\rangle\subseteq {\rm Im}(\Phi)$.

\par
  As stated above, we conclude that ${\rm Im}(\Phi)=\langle (x^N-a_0)^{p^r-k}\rangle$.
\hfill $\Box$

\vskip 3mm \par
   By the scalar multiplication defined by Equation (6), we see that $F_{p^m}[x]$-submodules
of the ring $F_{p^m}[x]/\langle x^{p^rN}-\vartheta\rangle$ are ideals of $F_{p^m}[x]/\langle x^{p^rN}-\vartheta\rangle$,
i.e., $\vartheta$-constacyclic codes over $F_{p^m}$ of length $p^rN$. Then by Lemma 4.2 and isomorphism theorems of module theory,
we conclude that $\Phi$ induces a bijection from the set of
$F_{p^m}[x]$-submodules
of $R_k[x]/\langle x^N-\lambda\rangle$ onto the set of ideals of $F_{p^m}[x]/\langle x^{p^rN}-\vartheta\rangle$ contained in $\langle (x^N-a_0)^{p^r-k}\rangle$. Precisely, we have the following corollary.

\vskip 3mm \noindent
  {\bf Corollary 4.3} \textit{Let $C\subseteq R_k[x]/\langle x^N-\lambda\rangle$. Then $C$ is an $F_{p^m}[x]$-submodule
of the ring $R_k[x]/\langle x^N-\lambda\rangle$ if and only if there is a unique
$\vartheta$-constacyclic code $\mathcal{C}$ over $F_{p^m}$ of length $p^rN$ contained in $\langle (x^N-a_0)^{p^r-k}\rangle$ such
that $\Phi(C)=\mathcal{C}$}.

\vskip 3mm \par
   Next, we determine the ideals of $F_{p^m}[x]/\langle x^{p^rN}-\vartheta\rangle$ contained in $\langle (x^N-a_0)^{p^r-k}\rangle$.
Since $F_{p^m}^{\times}$ is a multiplicative group
of order $p^m-1$ and $a_0\in F_{p^m}^{\times}$, there is a unique element $\theta\in F_{p^m}^{\times}$ such that $$\theta^{p^e}=a_0.$$
As ${\rm gcd}(p,n)=1$, there are pairwise coprime irreducible monic polynomials $f_1(x),f_2(x),\ldots,f_t(x)\in F_{p^m}[x]$
such that
$$x^n-\theta=f_1(x)f_2(x)\ldots f_t(x)$$
From this and by $\vartheta=a_0^{p^r}$, we deduce
\begin{eqnarray*}
x^{p^rN}-\vartheta&=&(x^N-a_0)^{p^r-k}(x^N-a_0)^k=(x^N-a_0)^{p^r-k}(x^n-\theta)^{p^ek}\\
&=&(x^N-a_0)^{p^r-k}f_1(x)^{p^ek}\ldots f_t(x)^{p^ek}.
\end{eqnarray*}
Since $F_{p^m}[x]$ is a principal ideal ring, by the classical ring theory we have the following lemma.

\vskip 3mm \noindent
   {\bf Lemma 4.4} \textit{All distinct $\vartheta$-constacyclic codes
over $F_{p^m}$ of length $p^rN$ contained in $\langle (x^N-a_0)^{p^r-k}\rangle$ are given by the following}:
$${\cal C}_{(l_1,\ldots,l_t)}=\left\langle (x^N-a_0)^{p^r-k}f_1(x)^{l_1}\ldots f_t(x)^{l_t}\right\rangle \ ({\rm mod} \ x^{p^rN}-\vartheta),$$
\textit{where $0\leq l_1,\ldots,l_t\leq p^ek$. Hence the number of such codes is equal to $(p^ek+1)^t$. Moreover, the number of
codewords in ${\cal C}_{(l_1,\ldots,l_t)}$ is equal to}
$$|{\cal C}_{(l_1,\ldots,l_t)}|=p^{m(Nk-\sum_{j=1}^tl_j{\rm deg}(f_j(x)))}.$$

\vskip 3mm \noindent
   {\bf Proof.} We only need to calculate the number of elements in ${\cal C}_{(l_1,\ldots,l_t)}$.
Since $(x^N-a_0)^{p^r-k}f_1(x)^{l_1}\ldots f_t(x)^{l_t}$ is a factor of $x^{p^rN}-\vartheta$ in $F_{p^m}[x]$ and
$${\rm deg}((x^N-a_0)^{p^r-k}f_1(x)^{l_1}\ldots f_t(x)^{l_t})=N(p^r-k)+\sum_{j=1}^tl_j{\rm deg}(f_j(x)),$$
as an ideal of $F_{p^m}[x]/\langle x^{p^rN}-\vartheta\rangle$ generated by $(x^N-a_0)^{p^r-k}f_1(x)^{l_1}\ldots f_t(x)^{l_t}$
we have
$|{\cal C}_{(l_1,\ldots,l_t)}|=(p^m)^{p^rN-\left(N(p^r-k)+\sum_{j=1}^tl_j{\rm deg}(f_j(x))\right)}=p^{m(Nk-\sum_{j=1}^tl_j{\rm deg}(f_j(x)))}.$
\hfill $\Box$

\vskip 3mm \par
   Finally, we determine all ideals of the ring $R_k[x]/\langle x^N-\lambda\rangle$.

\vskip 3mm \noindent
   {\bf Theorem 4.5} \textit{Using notations above, all distinct
$\lambda$-constacyclic codes over $R_k$ of length $N$ are given by the following}
$$C_{(l_1,\ldots,l_t)}=\langle f_1(x)^{l_1}\ldots f_t(x)^{l_t}\rangle \ ({\rm mod} \ x^N-\lambda), \ 0\leq l_1,\ldots,l_t\leq p^ek.$$
\textit{Moreover, we have $\Phi(C_{(l_1,\ldots,l_t)})=\mathcal{C}_{(l_1,\ldots,l_t)}$ and}
 $$|C_{(l_1,\ldots,l_t)}|=p^{m(Nk-\sum_{j=1}^tl_j{\rm deg}(f_j(x)))}, \ 0\leq l_1,\ldots,l_t\leq p^ek.$$

\par
   \textit{Hence the number of $\lambda$-constacyclic codes over $R_k$ of length $N$ is equal to $(p^ek+1)^t$.}

\vskip 3mm \noindent
   {\bf Proof.} In the ring $R_k[x]/\langle x^N-\lambda\rangle$, by Corollary 2.2(ii)
it follows that
$$(1,u,\ldots,u^{k-1})=\left(1,(x^N-a_0),(x^N-a_0)^2,\ldots,(x^N-a_0)^{k-1}\right)(Y^{-1}P_k)$$
where $Y^{-1}P_k\in {\rm M}_{k\times k}(F_{p^m})$, which implies that
$u=\sum_{i=0}^{k-1}b_i(x^N-a_0)^i$ for some $b_0,b_1,\ldots,b_{k-1}\in F_{p^m}$.

\par
  Let $C$ be any $F_{p^m}[x]$-submodule of $R_k[x]/\langle x^N-\lambda\rangle$. As
$\sum_{i=0}^{k-1}b_i(x^N-a_0)^i\in F_{p^m}[x]$, we conclude that
$uC=(\sum_{i=1}^{k-1}b_i(x^N-a_0)^i)C\subseteq C.$
Hence $C$ must be an ideal
of $R_k[x]/\langle x^N-\lambda\rangle$. From this and by Corollary 4.3, we deduce that the map $C\mapsto \Phi(C)$
is a bijection from the set of ideals in $R_k[x]/\langle x^N-\lambda\rangle$ onto
the set of ideal in $F_{p^m}[x]/\langle x^{p^rN}-\vartheta\rangle$ contained in $\langle (x^N-a_0)^{p^r-k}\rangle$.

\par
  Now, let $\mathcal{C}_{(l_1,\ldots,l_t)}=\langle (x^N-a_0)^{p^r-k}f_1(x)^{l_1}\ldots f_t(x)^{l_t}\rangle$
be any ideal of the ring $F_{p^m}[x]/\langle x^{p^rN}-\vartheta\rangle$ contained in $\langle (x^N-a_0)^{p^r-k}\rangle$,
where $0\leq l_1,\ldots,l_t\leq p^ek$. As a polynomial in $F_{p^m}[x]$,
$f_1(x)^{l_1}\ldots f_t(x)^{l_t}$ can be uniquely expressed as
\begin{equation}
f_1(x)^{l_1}\ldots f_t(x)^{l_t}=\sum_{i=0}^{p^r-1}h_i(x)(x^N-a_0)^i
\end{equation}
 for some $h_0(x),h_1(x),\ldots,h_{p^r-1}(x)\in
F_{p^m}[x]$ both having degrees less that $N$. Then by $(x^N-a_0)^{p^r}=x^{p^rN}-\vartheta=0$
in $F_{p^m}[x]/\langle x^{p^rN}-\vartheta\rangle$, we have
\begin{eqnarray*}
&&(x^N-a_0)^{p^r-k}f_1(x)^{l_1}\ldots f_t(x)^{l_t}\\
   &=&(x^N-a_0)^{p^r-k}\sum_{i=0}^{k-1}h_i(x)(x^N-a_0)^i\\
  &=&(\omega^{p^r-k},\omega^{p^r-k+1},\ldots,\omega^{p^r-1})H(1,x,x^2,\ldots,x^{N-1})^{{\rm tr}},
\end{eqnarray*}
where $\omega=x^N-a_0$ and $H\in {\rm M}_{k\times N}(F_{p^m})$ satisfying $H(1,x,x^2,\ldots,x^{N-1})^{{\rm tr}}$ $=
(h_0(x),h_1(x),\ldots,h_{k-1}(x))^{{\rm tr}}$. Now, we set
$$c(x)=(1,u,\ldots,u^{k-1})(P_k^{-1}H)(1,x,x^2,\ldots,x^{N-1})^{{\rm tr}}\in R_k[x]/\langle x^N-\lambda\rangle.$$
Then by Lemma 4.1, it follows that
\begin{eqnarray*}
\Phi(c(x))
&=&(\omega^{p^r-k},\omega^{p^r-k+1},\ldots,\omega^{p^r-1})(P_k(P_k^{-1}H))(1,x,x^2,\ldots,x^{N-1})^{{\rm tr}}\\
&=&(x^N-a_0)^{p^r-k}f_1(x)^{l_1}\ldots f_t(x)^{l_t},
\end{eqnarray*}
which implies that
\begin{eqnarray*}
\Phi(\langle c(x)\rangle)&=&\langle\Phi(c(x))\rangle\\
  &=&\left\langle(\omega^{p^r-k},\omega^{p^r-k+1},\ldots,\omega^{p^r-1})H(1,x,x^2,\ldots,x^{N-1})^{{\rm tr}}\right\rangle\\
  &=& \langle (x^N-a_0)^{p^r-k}f_1(x)^{l_1}\ldots f_t(x)^{l_t}\rangle\\
  &=&\mathcal{C}_{(l_1,\ldots,l_t)}.
\end{eqnarray*}
Since $\Phi$ induces a bijection from $\langle c(x)\rangle$ onto $\mathcal{C}_{(l_1,\ldots,l_t)}$,
 by Lemma 4.4 it follows that
$|\langle c(x)\rangle|=|\mathcal{C}_{(l_1,\ldots,l_t)}|=p^{m(Nk-\sum_{j=1}^tl_j{\rm deg}(f_j(x)))}$.

\par
  By Theorem 2.1 and Corollary 2.2(ii), we have $P_k^{-1}J_k=a(J_k)P_k^{-1}$
and $Y=I_k+\sum_{i=1}^{k-1}y_i^{(k)}J_k^i$ respectively, which imply
\begin{eqnarray*}
P_k^{-1}Y&=&P_k^{-1}\cdot I_k+\sum_{i=1}^{k-1}y_i^{(k)}(P_k^{-1}\cdot J_k^i)=I_k\cdot P_k^{-1}+\sum_{i=1}^{k-1}y_i^{(k)}a(J_k)^iP_k^{-1}\\
  &=&\left(I_k+\sum_{i=1}^{k-1}y_i^{(k)}a(J_k)^i\right)P_k^{-1}.
\end{eqnarray*}
As $(1,u,\ldots,u^{k-1})a(J_k)=a(u)(1,u,\ldots,u^{k-1})$ by Corollary 2.2(i), it follows that
$(1,u,\ldots,u^{k-1})a(J_k)^i=a(u)^i(1,u,\ldots,u^{k-1})$ for all $i$. Hence
\begin{eqnarray*}
(1,u,\ldots,u^{k-1})P_k^{-1}Y&=&(1,u,\ldots,u^{k-1})\left(I_k+\sum_{i=1}^{k-1}y_i^{(k)}a(J_k)^i\right)P_k^{-1}\\
&=&g(u)(1,u,\ldots,u^{k-1})P_k^{-1},
\end{eqnarray*}
where $g(u)=\left(1+\sum_{i=1}^{k-1}y_i^{(k)}a(u)^i\right)\in R_k$. Since $a(u)=u(1+\sum_{i=2}^{k-1}a_iu^{i-1})$
and $u^k=0$, we see that $g(u)\in R_k^{\times}$. We assume $h(u)=g(u)^{-1}\in R_k^{\times}$. Then
by $h(u)g(u)=1$ in $R_k$, we have
\begin{equation}
h(u)(1,u,\ldots,u^{k-1})P_k^{-1}Y=(1,u,\ldots,u^{k-1})P_k^{-1}.
\end{equation}

\par
   In the ring $R_k[x]/\langle x^N-\lambda\rangle$, by Corollary 2.2(ii) we have
$(1,\omega,\ldots,\omega^{k-1})=(1,u,\ldots,u^{k-1})P_k^{-1}Y$ and
$\omega^{k}=(u+\sum_{i=2}^{k-1}a_iu^i)=u^k(1+\sum_{i=2}^{k-1}a_iu^{i-1})^k=0$. From this and by Equation (7), we deduce that
\begin{eqnarray*}
f_1(x)^{l_1}\ldots f_t(x)^{l_t}&=&\sum_{i=0}^{p^r-1}h_i(x)(x^N-a_0)^i=\sum_{i=0}^{k-1}h_i(x)\omega^i\\
  &=&(1,\omega,\ldots,\omega^{k-1})(h_0(x),h_1(x),\ldots,h_{k-1}(x))^{{\rm tr}}\\
  &=&(1,u,\ldots,u^{k-1})P_k^{-1}Y\cdot H(1,x,x^2,\ldots,x^{N-1})^{{\rm tr}}
\end{eqnarray*}
as an element of $R_k[x]/\langle x^N-\lambda\rangle$. Moreover, by Equation (8) it follows that
$$h(u)f_1(x)^{l_1}\ldots f_t(x)^{l_t}=(1,u,\ldots,u^{k-1})P_k^{-1}\cdot H(1,x,x^2,\ldots,x^{N-1})^{{\rm tr}}.$$
Therefore, by Lemma 4.1 it follows that
\begin{eqnarray*}
&&\Phi(h(u)f_1(x)^{l_1}\ldots f_t(x)^{l_t})\\
&=&(\omega^{p^r-k},\omega^{p^r-k+1},\ldots,\omega^{p^r-1})
   \left(P_k(P_k^{-1}H)\right)(1,x,x^2,\ldots,x^{N-1})^{{\rm tr}}\\
&=&(\omega^{p^r-k},\omega^{p^r-k+1},\ldots,\omega^{p^r-1})
   H(1,x,x^2,\ldots,x^{N-1})^{{\rm tr}}\\
&=&(x^N-a_0)^{p^r-k}f_1(x)^{l_1}\ldots f_t(x)^{l_t}\in F_{p^m}[x]/\langle x^{p^rN}-\vartheta\rangle,
\end{eqnarray*}
which implies that $\Phi(c(x))=\Phi(h(u)f_1(x)^{l_1}\ldots f_t(x)^{l_t})$. Since $\Phi$ is injective, we
have $c(x)=h(u)f_1(x)^{l_1}\ldots f_t(x)^{l_t}$. As $h(u)$ is invertible in $R_k[x]/\langle x^N-\lambda\rangle$, it follows that
$\langle c(x)\rangle=\langle f_1(x)^{l_1}\ldots f_t(x)^{l_t}\rangle$ as an ideal of $R_k[x]/\langle x^N-\lambda\rangle$. Hence
$$\mathcal{C}_{(l_1,\ldots,l_i)}=\Phi(\langle c(x)\rangle)=\Phi(\langle f_1(x)^{l_1}\ldots f_t(x)^{l_t}\rangle)
=\Phi(C_{(l_1,\ldots,l_i)}).$$

\par
  As stated above, we conclude that all distinct ideals of
$R_k[x]/\langle x^N-\lambda\rangle$ are given by: $C_{(l_1,\ldots,l_t)}=\langle f_1(x)^{l_1}\ldots f_t(x)^{l_t}\rangle$, $0\leq l_1,\ldots,l_t\leq p^ek$.
\hfill $\Box$

\vskip 3mm \noindent
   {\bf Remark} For any integers $0\leq l_1,\ldots,l_t\leq p^ek$,
by Theorems 3.6 and 4.5, we see that the Lee weight (distance) distribution of the $\lambda$-constacyclic code
$C_{(l_1,\ldots,l_t)}=\langle f_1(x)^{l_1}\ldots f_t(x)^{l_t}\rangle$ over $R_k$ of length $N$ is the same as the Hamming
weight (distance) distribution of the $\vartheta$-constacyclic code
$\mathcal{C}_{(l_1,\ldots,l_t)}=\langle (x^N-a_0)^{p^r-k}f_1(x)^{l_1}\ldots f_t(x)^{l_t}\rangle$ over $F_{p^m}$ of length $p^rN$.

%%%%%%%%%%%%%%%%%%%%%%%%%%%%%%%%%%%%%%%%%%%%%%%%%%%%%%%%%%%%%%%%%%%%%%%%%%%%%%%%%%%%%%%%%%

%%%%%%%%%%%%%%%%%%%%%%%%%%%%%%%%%%%%%%%%%%%%%%%%%%%%%%%%%%%%%%%%%%%%%%%%%%%%%%%%%%%%%%%%%%%%%%
\section{Examples} \label{}
\noindent
  \par  First we define a Gray map $\varphi$ from $R_k$ to $F_{p^m}^{p^r}$ by Definition 3.1. Then we obtain the Lee weight of each element in $R_k$ by computing the Hamming weight of its corresponding Gray image in $F_{p^m}^{p^r}$. By constructing all $\lambda$-constacyclic codes over $R_k$ of length $N$, we determine all $\vartheta$- constacyclic codes over $F_{p^m}$. Finally, we obtain optimal $\vartheta$-constacyclic codes over $F_{p^m}$.

 \vskip 3mm \noindent{\bf Example 5.1} Let $p=3$, $k=3$, $m=1$, $r=1$, and $\lambda=2+u+2u^2$. We consider $(2+u+2u^2)$-constacyclic codes over $R_3=F_3[u]/\langle u^3\rangle$. Then we define a Gray map $\varphi$ from $R_3$ to $F_3^3$ by Definition 3.1, and obtain the Lee weight of each element in $R_3$ by computing the Hamming weight of its corresponding Gray image in $F_3^3$.
  \par The Gray map $\varphi_1$ from $R_3$ to $F_3^3$ and Lee weight of $27$ elements of $R_3$ list as the following table, where $\alpha$ is the element in $R_3$, $\varphi_1(\alpha)$ is the Gray image of $\alpha$ in $F_3^3$, and ${\rm wt}_L(\alpha)$ is the Lee weight of $\alpha$ in $R_3$.
  {\small \begin{center}
\begin{tabular}{lll|lll}\hline
$\alpha$  & $\varphi_1(\alpha)$ & ${\rm wt}_L(\alpha)$ &$\alpha$  & $\varphi_1(\alpha)$ & ${\rm wt}_L(\alpha)$   \\ \hline
$0$  & $(0,0,0)$ & $0$ & $2u^2+u+1$  & $(0,1,2)$ & $2$  \\
$u^2$  & $(1,2,1)$ & $3$ & $2u+1$  & $(2,1,0)$ & $2$  \\
$2u^2$  & $(2,1,2)$ & $3$ & $u^2+2u+1$  & $(0,0,1)$ & $1$  \\
$u$  & $(1,1,0)$ & $2$ & $2u^2+2u+1$  & $(1,2,2)$ & $3$  \\
$u^2+u$  & $(2,0,1)$ & $2$ & $2$  & $(0,1,0)$ & $1$  \\
$2u^2+u$  & $(0,2,2)$ & $2$ & $u^2+2$  & $(1,0,1)$ & $2$  \\
$2u$  & $(2,2,0)$ & $2$ & $2u^2+2$  & $(2,2,2)$ & $3$  \\
$u^2+2u$  & $(0,1,1)$ & $2$ & $u+2$  & $(1,2,0)$ & $2$  \\
$2u^2+2u$  & $(1,0,2)$ & $2$ & $u^2+u+2$  & $(2,1,1)$ & $3$  \\
$1$  & $(0,2,0)$ & $1$ & $2u^2+u+2$  & $(0,2,2)$ & $1$  \\
$u^2+1$  & $(1,1,1)$ & $3$ & $2u+2$  & $(2,0,0)$ & $1$  \\
$2u^2+1$  & $(2,0,2)$ & $2$ & $u^2+2u+2$  & $(0,2,1)$ & $2$  \\
$u+1$  & $(1,0,0)$ & $1$ & $2u^2+2u+2$  & $(1,1,2)$ & $3$  \\
$u^2+u+1$  & $(2,2,1)$ & $3$ & $$  & $$ & $$  \\
\hline
\end{tabular}
\end{center}}

 \vskip 3mm In this example, $a_0=2$, $p=3$, and $r=1$, so $\vartheta=a_0^{p^r}=2^{3^1}=2 \ ({\rm mod}\ 3)$. We take the length of codes $N$ as 4, 5, and 6.
  \par
  $\bullet$ $N=4$, $N=p^en$ and $\theta^{p^e}=a_0$, so $e=0$, $n=4$, and $\theta=2$. Then in $F_3[x]$, $x^n-\theta=x^4-2=f_1f_2$, where $f_1=x^2+x+2$, $f_2=x^2+2x+2$.
  \par
  $\bullet$ $N=5$, $N=p^en$ and $\theta^{p^e}=a_0$, so $e=0$, $n=5$, and $\theta=2$. Then in $F_3[x]$, $x^n-\theta=x^5-2=g_1g_2$, where $g_1=x+1$, $g_2=x^4+2x^3+x^2+x+2$.
  \par
  $\bullet$ $N=6$, $N=p^en$ and $\theta^{p^e}=a_0$, so $e=1$, $n=2$, and $\theta=2$. Then in $F_3[x]$, $x^n-\theta=x^2-2=h_1$, where $h_1=x^2+1$.
 \par
 Then we construct all $(2+u+2u^2)$-constacyclic codes over $R_3$ of length 4, 5, 6 respectively by Theorem 4.5, and obtain the corresponding $2$-constacyclic codes over $F_3$ by distance-preserving map $\Phi$ defined in Theorem 3.6. We list optimal $2$-constacyclic codes over $F_3$ in the following table. In this table,$C=\langle g \rangle$ is a $(2+u+2u^2+4u^3)$-constacyclic code over $R_4$,  where $g$ is the generator polynomial of $C$, $N$ is the length of $C$, and $\mathcal{C}=\Phi(C)$ is the corresponding optimal $2$-constacyclic code over $F_3$ with parameter $[n,k,d]_3$ (see [6]).
  \vskip 3mm
   {\small \begin{center}
  \begin{tabular}{lll|lll}\hline
$N$ & $C$ & $\mathcal{C}$ & $N$  & $C$ & $\mathcal{C}$   \\ \hline
4  & $\langle f_2\rangle$ & $[12,10,2]_3$   & 4  & $\langle f_1^3f_2^2\rangle$& $[12,2,9]_3$ \\
4  & $\langle f_1\rangle$ & $[12,10,2]_3$  &5  & $\langle g_1\rangle$ & $[15,14,2]_3$\\
4  & $\langle f_2^2\rangle$ & $[12,8,3]_3$ &5  & $\langle g_1^2\rangle$ &$[15,13,2]_3$  \\
4  & $\langle f_1^2\rangle$ & $[12,8,3]_3$ & 5  & $\langle g_1^3\rangle$ & $[15,12,2]_3$  \\
4  & $\langle f_1f_2^3\rangle$ & $[12,4,6]_3$ & 5 & $\langle g_2\rangle$ & $[15,11,3]_3$    \\
4  & $\langle f_1^3f_2\rangle$  & $[12,4,6]_3$ & 5  & $\langle g_1^2g_2\rangle$& $[15,9,4]_3$ \\
4  & $\langle f_1^2f_2^3\rangle$ &$[12,2,9]_3$ &6 & $\langle h_1\rangle$ & $[18,16,2]_3$ \\
\hline
\end{tabular}
\end{center}}
    \vskip 3mm \noindent
\vskip 3mm \noindent{\bf Example 5.2} Let $p=5$, $k=4$, $m=1$, $r=1$, and $\lambda=2+u+2u^2+4u^3$. We consider $(2+u+2u^2+4u^3)$-constacyclic codes over $R_4=F_5[u]/\langle u^4\rangle$. Similar to Example 5.1, we obtain the Gray map $\varphi$ from $R_4$ to $F_5^5$ and Lee weight of each element in $R_4$. To save the space, we do not list the table.
\par In this example, $a_0=2$, $p=5$, and $r=1$, so $\vartheta=a_0^{p^r}=2^{5^1}=2 \ ({\rm mod}\ 5)$. We take the length of codes $N$ as 3 and 5.
 \par
  $\bullet$ $N=3$, $N=p^en$ and $\theta^{p^e}=a_0$, so $e=0$, $n=3$, and $\theta=2$. Then in $F_5[x]$, $x^n-\theta=x^3-2=f_1f_2$, where $f_1=x+2$, $f_2=x^2+3x+4$.
  \par
  $\bullet$ $N=5$, $N=p^en$ and $\theta^{p^e}=a_0$, so $e=1$, $n=1$, and $\theta=2$. Then in $F_5[x]$, $x^n-\theta=x-2=g_1$, where $g_1=x+3$.
  \par Then we construct all $(2+u+2u^2+4u^3)$-constacyclic codes over $R_4$ of length 3 and 5 respectively by Theorem 4.5, and obtain the corresponding $2$-constacyclic codes over $F_5$ by the distance-preserving map $\Phi$ defined in Theorem 3.6. We list optimal $2$-constacyclic codes over $F_5$ in the following table. In this table, $C=\langle g \rangle$ is a $(2+u+2u^2+4u^3)$-constacyclic code over $R_4$,  where $g$ is the generator polynomial of $C$, $N$ is the length of $C$, and $\mathcal{C}=\Phi(C)$ is the corresponding optimal $2$-constacyclic code over $F_5$ with parameter $[n,k,d]_5$ (see [6]).
    \vskip 3mm
   {\small \begin{center}
   \begin{tabular}{lll}\hline
$N$ & $C$ & $\mathcal{C}$ \\ \hline
3 & $f_1^2f_2^4$ & $[15,2,12]_5$\\
3  & $f_1^2$ & $[15,10,4]_5$   \\
 5 & $g_1^{19}$ &$[25,2,20]_5$ \\
\hline
\end{tabular}
\end{center}}
\vskip 3mm \noindent
   {\bf Remark} One can obtain more optimal $\vartheta$-constacyclic codes over $F_{p^m}$ of length $p^rN$ with other setting of $p$, $k$, $m$, $r$, $N$, and $\lambda$.
%%%%%%%%%%%%%%%%%%%%%%%%%%%%%%%%%%%%%%%%%%%%%%%%%%%%%%%%%%%%%%%%%%%%%%%%%%%%%%%%%%%%%%%%%%

%%%%%%%%%%%%%%%%%%%%%%%%%%%%%%%%%%%%%%%%%%%%%%%%%%%%%%%%%%%%%%%%%%%%%%%%%%%%%%%%%%%%%%%%%%%%%%

\vskip 3mm \noindent {\bf Acknowledgments}
 Part of this work was done when Yonglin Cao was visiting Chern Institute of Mathematics, Nankai University, Tianjin, China. Yonglin Cao would like to thank the institution for the kind hospitality. This research is
supported in part by the National Natural Science Foundation of
China (Grant Nos. 11671235, 11471255, 61171082) and the National Key Basic Research Program of China (Grant No. 2013CB834204) .

%\bibliographystyle{ieicetr}% bib style
%\bibliography{}% your bib database

\end{document}